Vladislav G. Polnikov

# FEATURES OF AIR FLOW

# IN THE TROUGH-CREST ZONE OF WIND WAVES

A.M. Obukhov Institute of Atmospheric Physics of Russian Academy of Sciences

Pyzhevskii lane 3, 119017, Moscow, Russia**,**

E-mail: polnikov@mail.ru

Tel: +7-916-3376728

Fax: +7-495-9531652

**Abstract**

Vertical profiles for mean wind, standard deviations of velocity fluctuations, and wave-induced part of the momentum flux over a wavy fluid surface are calculated in the Cartesian coordinates on the basis of recent numerical results by Chalikov and Rainchik. Besides, calculations of spectra for surface elevation and wave-induced velocity components at this surface are carried out. Unlike the profiles typical to the near-wall turbulence, the calculated wind-velocity profiles deviate significantly from the logarithmic law throughout the entire trough-crest zone of wind waves, and the wave-induced part of the momentum flux changes its sign in close vicinity of the mean surface level. Vertical scales of the profiles features are estimated. Spectral analysis of the water-surface oscillations and wind-velocity components suggests nonlinear and strongly anisotropic dynamics of the system under consideration. Points of application of the results obtained are discussed.

Key words: atmospheric boundary layer, surface elevation, wind waves, trough-crest zone, mean wind profile, spectra.

1. **Introduction**

Studying the dynamics of interaction between wind flow and wavy fluid surface is a very complicated scientific and technical problem, as far as the flow, wave and turbulent motions exist simultaneously in this system, differing significantly by scales of their variability (Sulliwan et al 2000; Chalikov and Rainchik 2010; Cavalery et al. 2007; Polnikov 2009a). Therefore, due to technical reasons, experimental observations of fine details of the interface dynamics are significantly restricted.

At present, owing to computer technique development, some kinds of observations of the values said above could be replaced by numerical simulations. A leading position in this field belongs to D.V. Chalikov. He started this study in the last third of the former century and is permanently continuing it up to date (for example, Chalikov 1978, 1986; Chalikov and Sheinin 1998; Chalikov and Rainchik 2010). Last two decades his efforts have been accompanied with numerous works by other authors, carried out on the basis of different numerical approaches (for references see Sulliwan et al 2000; Chalikov and Rainchik 2010).

Not dwelling on mathematical grounds of these papers, notice only that in the last paper (Chalikov and Rainchik 2010), in the frame of $K$-$\varepsilon$ turbulence closure technique, a large series of simulations related to calculation of numerous characteristics of air flow over wind waves was executed. Using a specially introduced curvilinear coordinate system, the authors have calculated a two-dimensional wind velocity field in the air boundary layer over random wavy surface of water. In these coordinates they estimated a lot of characteristics of the system, including vertical profiles of all constituents for the momentum flux from wind to waves. Besides, the rate of energy supply from wind to waves was calculated and parameterized, and the dependence of drag coefficient on integral parameters of the system was established. However, the point of presentation of the wind profile and the momentum flux in the Cartesian coordinate system was not considered in the paper referred. Present paper is devoted to this particular issue. The key point of our work is to extend an analysis of some characteristics of the system to the whole interface domain including the zone of wave troughs, located below the mean surface level.

**2. Problem formulation**

Formulation of the problem is the following. Basing on the data of wind field in the air boundary layer over waves, having a vertical scale less then dominant wave length, we would like to calculate the mean wind profile and the wave-induced part of the momentum flux in the Cartesian coordinates (i.e. the coordinate system of a stationary observer) for the whole simulation domain including the trough-crest zone of wind waves. In our knowledge, calculations of such a kind, including the zone below the mean surface level, were not presented in the world scientific literature yet.

A meaning of these calculations, besides direct scientific interest, is conditioned by numerous practical tasks which are usually solved in the Cartesian coordinates. For example, they could be estimations of some air-sea interaction values, based on bulk-formulas, including large scales exchange of heat, gas, mechanical energy, and so on.

From scientific point of view, the deviations of profiles over wavy surface from the profiles corresponding to the case of hard-wall turbulence are of the most interest just in the Cartesian coordinates, as far as in these coordinates the deviations should be the most evident. Owing to this circumstance, calculations of these deviations extend our possibilities for experimental checking a quality of numerical models, results of which are used in the proper comparison.

**3. Input data**

Initial data used in our calculations are two-dimensional arrays of the following values: the horizontal and vertical Cartesian coordinates, $x(i,j)$, $z(i,j)$, and the horizontal and vertical Cartesian wind velocity components, $u(i,j)$, $w(i,j)$. All these values have been calculated by Chalikov and Rainchik (2010) in the curvilinear coordinate system for a lot of time series and states of wind-wave situations. A small part of these data were kindly provided by Chalikov to the author for the present calculations having a pilot feature.

The two-dimensional curvilinear grid, $\{\zeta(i,j), \xi(i,j)\}$, was constructed in such a kind that the "horizontal" grid lines $\zeta(i,j)$ are properly following oscillations of the fluid surface, and the "vertical" grid lines at each grid node are orthogonal to the "horizontal" ones (for details, see the original paper). The number of the "horizontal" and "vertical" nodes was $i = 512$ and $j = 61$, respectively. Herewith, a set of "vertical" curvilinear coordinates $\xi(i,j)$ with index $j = 11$ was corresponding to the fluid surface.

The dimensional horizontal size of the calculation grid was given by an implicit value $D$ which was in dimensionless units equals to value $2\pi$ accepted for convenience of stating the periodical boundary conditions. All dimensional physical values were normalized by using the unity of the length, $L = D/2\pi$, and value of the gravity acceleration, $g$. Specification of value $D$ allows to recalculate all dimensionless characteristics of the numerical experiment into corresponding dimensional values, if it needs to specify any certain wind-wave situation. In the case considered here, the data correspond to the inverse wave age $A = U_{max} / C_p$ equals approximately to 2 (here, $U_{max}$ is the wind speed at the upper boundary of the modelling numerical domain, and $C_p$ is the phase speed of dominant wave).

### 4. Methodic of calculations

Methodic of our calculations has the following steps.

First. By means of averaging (over horizontal index *i*) each "vertical" curvilinear coordinate line $\xi(i,j)$, fixed Cartesian horizon $ZD(j)$ was estimated for each vertical index *j*. This is necessary to provide a set of Cartesian vertical coordinates $\{ZD(j)\}$, for which the mean profile will be calculated for any physical variable.

Second. By means of making a loop over vertical indexes *j*, for each horizontal index *i* of curvilinear grid and each velocity component *u* and *w*, their belonging to certain fixed horizon

*ZD(j)* was established. Herewith, a vertical location of fluid surface *ξ(i,11)* was especially controlled in system of Cartesian vertical coordinates *{ZD(j)}*. By this manner a set of proper one-dimensional arrays for physical values was defined at each fixed Cartesian horizon. Each of these variables was averaged over the horizontal index *i* and over numerous time series provided by the initial numerical simulations.

Third. The vertical profiles were estimated for the following values:

a) Mean horizontal speed  $<u> = U(z)$ ;

b) Mean vertical speed  $<w> = W(z)$;

c) Wave-induced vertical momentum flux  $\tau_w = <(u - U)(w - W)> = <uw>(z)$. [1]

Here, the brackets $<...>$ means averaging over index *i* and time series, *U* and *W* are the mean values of the proper velocity component, and momentum flux $\tau_w$ is corresponding to the wave-induced fluctuations of the wind flow, as far as the turbulent oscillations are not calculated in terms of *K-ε* turbulence closure technique used in the original work (Chalikov and Rainchik 2010).

Additionally, the horizontal wave number spectra were calculated for the surface elevation, $S_{\xi(11)}(k)$, and for the wind velocity components taken at the level of fluid surface, $S_{u(11)}(k)$, $S_{w(11)}(k)$.

For completeness of statistical analysis, the following vertical profiles were calculated:

$$D_u(z) = \left(<(u - U(z))^2>\right)^{1/2}, \tag{1}$$

$$D_w(z) = \left(<w^2>\right)^{1/2}, \tag{2}$$

$$E(z) = <\left((u - U(z))^2 + w^2\right)> . \tag{3}$$

Here, $D_u(z)$ and $D_w(z)$ are the standard deviations of the proper wind-velocity components, and $E(z)$ is the kinetic energy of their fluctuations.

---

[1] Hereafter in the definitions of mean values, it is accepted that the "theoretical mean value" of vertical velocity component *W* is equal to zero at each horizon.

The spectra are needed to determine the dominant wave scale and to control the statistical structure of the fields under consideration. The last three profiles represent the vertical distribution of standard deviations and energy of wave-induced velocity fluctuations.

**5. Results of calculations**

Calculation results for items a)-c) said above are presented in Figs. 1-3, respectively. The following features of the profiles are seen from these figures.

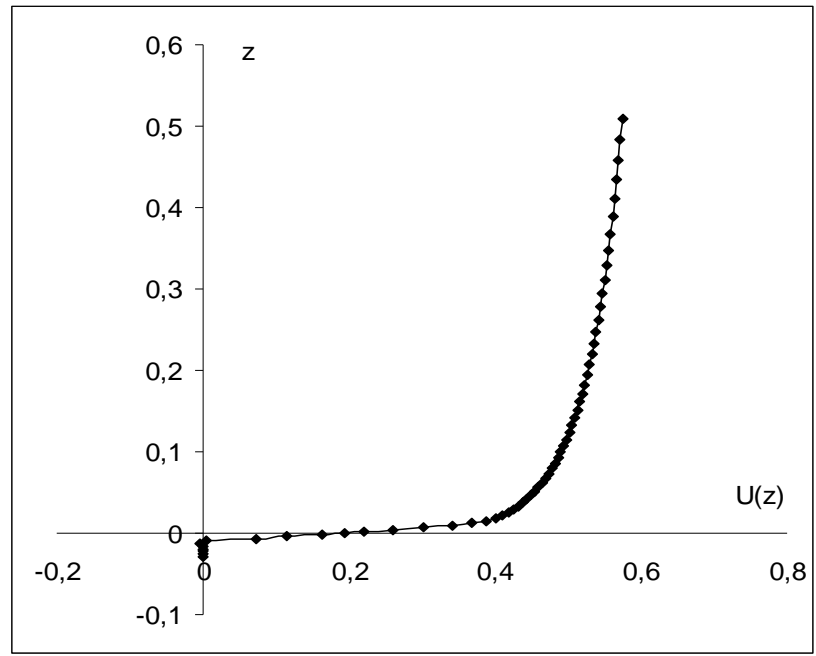

Fig. 1. Profile of mean horizontal wind component $U(z)$.

5.1. The upper part of profile $U(z)$, as expected, corresponds very well to the logarithmic law, i.e. $U(z) \propto \ln z$ (the semi-logarithmic presentation is given in Fig. 4 below). But, below the horizon corresponding to index $j=17$, profile $U(z)$ depends practically linearly on z, being extended down to the horizon with index $j=7$, located far lower the mean surface level $ZD(11) \cong 0$. These indexes correspond to the following fixed horizons: $ZD(7) = -9.79 \cdot 10^{-3}$, $ZD(17) = 1.54 \cdot 10^{-2}$.

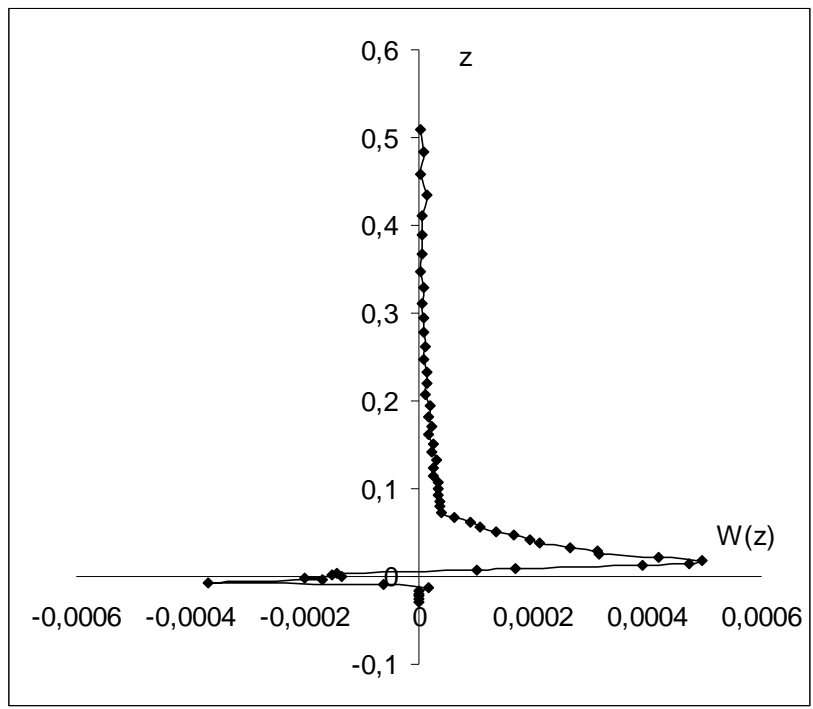

Fig. 2. Profile of mean vertical velocity *W(z)*.

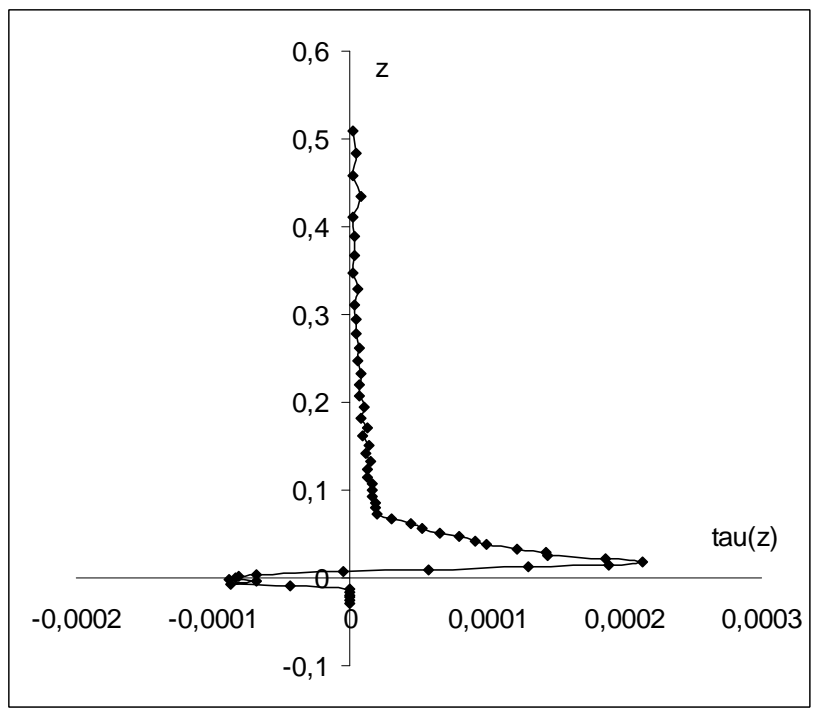

Fig. 3. Profile of wave-induced part of momentum flux $\tau_w(z)$.

Thus, we can state that the mean wind speed exists below the mean surface level, and the lower part of wind profile has a linear dependence on vertical coordinate far higher over this level.

Moreover, if we take into account that the mean wind speed at the upper boundary of numerical domain is given by the value $U(61) = U_{max} = 0.57$ [2], from Fig.1 we find that the mean wind at the point of the log-profile change has value $U(17) = 0.39$ which is remarkably greater than the phase speed of dominant wave, as far as $U(17)/U(61) \approx 0.68$ is grater than $C_p/U_{max} \cong 0.5$. This fact changes radically our understanding the wind profile formation over waves.

5.2. The mean value of the vertical velocity component $W$ is quite small along the whole profile (Fig. 2). Therefore, in the limits of errors of statistical averaging, from Fig. 2 one could state that estimations $W \cong 0$ is correct at all horizons, including the trough-crest zone of waves.

However, in the vicinity of point $j$=17, modeling profile $W(z)$ demonstrates a remarkable change resulting in negative values below the mean surface level ($j$=11). It seems that this result is not an error, as far as it appeared in each our calculations. Here we only can state that the feature revealed needs further intently investigation.

5.3. The profile of wave-induced part of momentum flux $\tau_w(z)$ has explicitly non-monotonic behavior expressed by a presence of maximum value at horizon $j$=17 and a sing change near the mean surface level with $j$=11 (Fig. 3). The maximum value is quite reasonable and equals to $\tau_{w,max} \cong 2.3 \cdot 10^{-4}$. [3]

The negative extremum of $\tau_w(z)$ has a value of the order of $1 \cdot 10^{-4}$, and this extremum is localized in close vicinity of the mean surface level. This result is explicitly out of averaging errors what allows saying about existence of the "sign inversion" effect as a new feature of profile $\tau_w(z)$, which was never published in literature yet.

---

[2] This value of $U_{max}$ is provided by the initial calculations.

[3] Here we should mention that according to more full estimations made in Chalikov and Rainchik (2010), $\tau_w$ contributes 20% of total momentum flux which additionally includes the turbulent part, and the total flux does not depend on vertical variable z, as it should be in a stationary turbulent shear flow near a wall (Monin and Yaglom 1971).

Presence of the sign inversion effect is apparently provided by the sign change in vertical profile *W(z)*, mentioned above (Fig. 2). Direct physical explanation of the effect is not so evident at present. Possibly, this fact is provided by the inverse action of troughs going upward quicker than downward. There are some other ideas of explanation, but before declaration of them, it needs additional theoretical justification of the effect in future. In any case, detailed study of this phenomenon, attracting different experimental and numerical techniques, is of great scientific interest.

5.4. Another point of our study is clarification a role of choice of reference level accepted for the profile calculation.[4] To solve this point, calculations of the mean wind profiles were executed under the following restrictions:

(a) Profile over troughs, only;

(b) Profile over the mean level, only;

(c) Profile over the level corresponding to the standard deviation of surface wave.

In case (a), the proper horizontal averaging *u(i,j)* was made for points where condition $\xi(i,11) < ZD(11)$ is met, only. In case (b), the same was done under condition $\xi(i,11) > ZD(11)$; and in case (c), the profile was found under condition $\xi(i,11) > DI$, where *DI* is the surface elevation standard deviation defined by the formula

$$DI = < \left(\xi(i,11) - ZD(11)\right)^2 >^{1/2} \qquad . \qquad (4)$$

In our calculation $DI = 4.77 \cdot 10^{-3}$.

Results of such a kind calculations are presented in Fig. 4 in the semi-logarithmic coordinates for more clarity.

---

[4] This point was mentioned in our previous paper (Volkov, Polnikov, Pogarkii 2007).

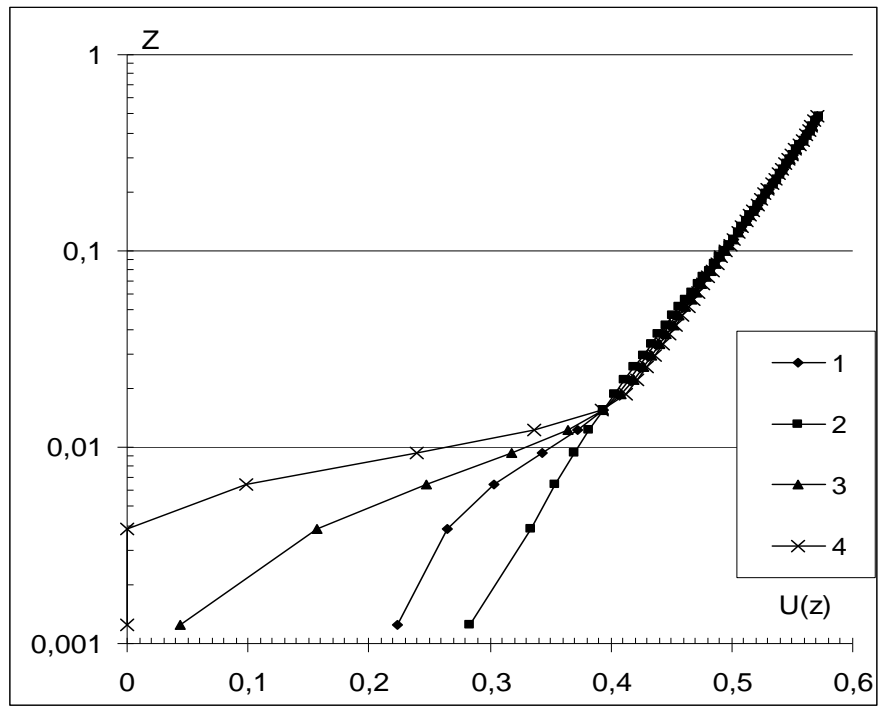

Fig. 4. Profiles of *U(z)* for different reference levels:
1- Total averaging (corresponding to Fig. 1), 2- case (a): averaging over troughs;
3-case (b): averaging over crests; 4- case (c): averaging over level *z=DI*.

As seen from Fig. 4, in the space domain above horizon $ZD(17)$, all the profiles coincide practically each to another and rigorously correspond to the logarithmic law (hereafter, log-law). This is a quite expected result, as we have mentioned earlier. But, below horizon $ZD(17)$ which is three times greater than the standard deviation of wavy surface: $ZD(17)/DI \geq 3$, the peculiarities of the mean wind profile, described earlier in details, become apparent. Herewith, in case (a), corresponding to the profile over troughs, the log-law for *U(z)* continues below the mean surface level (curve 2 in Fig. 4). This profile becomes linearly dependent on z from level $ZD(9) < 0$, resulting in U = 0 at level $ZD(7) = -9.79 \cdot 10^{-3}$.

In case (b) (curve 3 in Fig. 4), the linear falling-law for *U(z)* becomes steeper, and the mean wind is reaching zero value near horizon $ZD(11)$. This result is in full accordance with the formulation of task (b) [5].

---

[5] Analogous result was presented in Sulliwan et al (2000), when they considered the profile corresponding to case (b).

In the last case (c) (curve 4), the linear part of profile is going very steeply, getting zero value of the mean wind at level $z \approx DI$, as it should be in accordance with this task of calculations.

Herewith, it is important to note that in each case, the profiles are changing their dependence on the vertical coordinate at the same horizon $ZD(17) = 1.54 \cdot 10^{-2}$, which is three times above the standard deviation of wavy surface. This is a rather surprising result which needs its understanding further. Notice, additionally, that in the semi-log coordinates, the linear parts of profiles are not shown totally due to technical reasons. We only could add here that in the linear coordinates, the profiles have behavior similar to one shown in Fig.1.

5.5. Finally, the profiles of velocity components standard deviations and of velocity fluctuation energy are shown in Fig. 5. From this figure it is evident a quasi-exponential falling feature for standard deviations of the wave-induced velocity components in both directions from the mean surface level. Designations in the right-hand-side of Fig. 5 show that the intensity of fluctuations for horizontal component exceeds one for vertical component more than in one order. Herewith, the value $D_w(17) \cong 2 \cdot 10^{-2}$ at horizon $ZD(17) = 1.54 \cdot 10^{-2}$ is more than one order greater the mean value of vertical component at this horizon, $W(17) \cong 5 \cdot 10^{-4}$. This interesting fact emphasizes uncertainty of the latter value and stimulates necessity to check it in further investigations.

## 6. Dimensional quantities

To finish our work it needs to estimate dimensional quantities for characteristic values of the system under consideration. To do it, specify the horizontal size of the numerical domain, accepting, for example, value $D = 1000$m. Then, the dimensional units for length, [$L$], and speed, [$C$], are as follows

$$[L] = D/2\pi \approx 160\text{m} \quad \text{and} \quad [C] = \sqrt{Lg} \approx 39.6 \text{ m/s} . \tag{5}$$

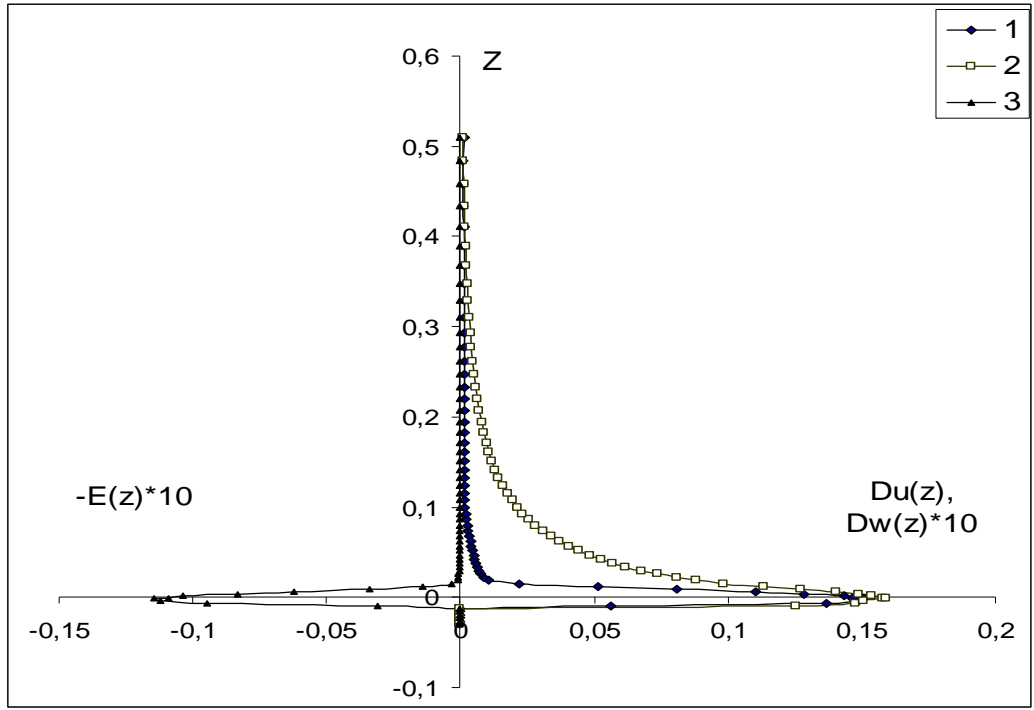

Fig. 5. Profiles for quadratic forms:

1-horizontal component s.d. $D_u(z)$; 2-vertical component s.d multiplied by 10, $D_w(z)*10$;

3-kinetic energy of velocity fluctuations multiplied by -10, $-E(z)*10$.

According to (5), one finds the following dimensional characteristic quantities:

$ZD(7) \approx$ -1.57m,

$ZD(17) \approx$ 2.46m,

$DI \approx$ 0.76m,

and the maximal wind speed,

$$U_{max} = U(61) \cdot [C] = 0.57\sqrt{Lg} \approx 22.6 \text{ m/s}. \qquad (6)$$

To compare the found estimation for $U_{max}$ with the phase speed of dominant waves $C_p$, one should calculate the peak parameters of the wave spectrum, as far as just they are characteristics of dominant waves. To do it, one needs to take into account that the horizontal space step is $\Delta x = D/512$, and the proper Niquist wave number is $k_N = 2\pi/2\Delta x = 2\pi*256/D$.[6] The said allows treating all parameters of space spectra S(k) calculated for any variable of the system. The proper spectra for surface elevation, $S_{\xi(11)}(k)$, and wave-induced fluctuation of horizontal wind velocity, $S_{u(11)}(k)$, are shown in Fig. 6 in conventional units and in the double logarithmic coordinates, calculated with 256 equidistant points in the whole band for wave numbers: $[0, k_N]$.

---

[6] These quantities are given with an error of the order of 1-2%, defined by the mean slope of the wave field.

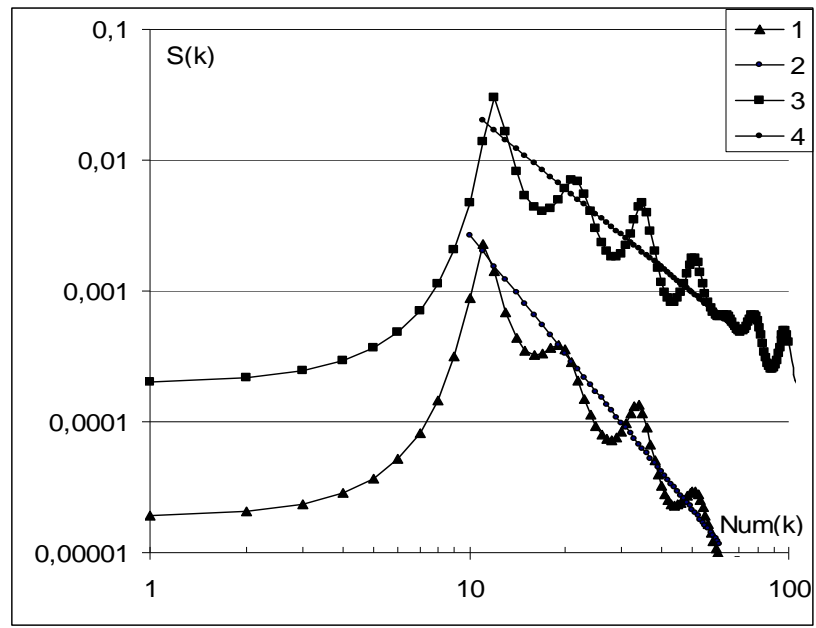

Fig. 6. Horizontal wave number spectra:
1- spectrum of surface elevation $S_{\xi(11)}(k)$; 2- reference line $const*k^{-3}$;
3-spectrum of horizontal wind component at the surface $S_{u(11)}(k)$; 4-reference line $const*k^{-2}$.

Analysis of Fig. 6 shows the following.

First. The maximum of spectrum $S_{\xi(11)}(k)$ corresponds to the 11$^{th}$ point on the wave number scale said above. Thus, for a peak wave number one has the following estimation

$$k_p = (11/256)\Delta k \approx 0.069 \text{ m}^{-1},$$

what, with using dispersion relation $\omega = (gk)^{1/2}$, gives the estimation

$$C_p = \omega_p / k_p \approx 11.9 \text{ m/s} \qquad . \tag{7}$$

Ratio (7) results in estimation $U_{max}/C_p \approx 1.9$ confirming the initial statement about a wave age of wind-wave situation under consideration.

Second. It is interesting to note that the falling law for the tail part of spectrum has the form

$S_{\xi(11)}(k) \propto k^{-3}$ which is very well corresponding to the initially accepted in Chalikov and Rainchik (2010) spectral shape having, in the high frequency domain, Phillip's falling law $S_{\xi(11)}(\omega) \propto \omega^{-5}$.

Along with the said, it is seen in Fig. 6 several excellently expressed peaks of high harmonics testifying to the nonlinear feature of interface dynamics. Interesting to note that the modeling surface elevation spectrum has a shape resembling very much a real wind-wave spectrum represented in paper by the author (Polnikov and Fedotov 1983). Detailed discussion of these points is out of this paper[7]. Here we would like only to emphasize that the similarity mentioned testifies to a high quality of the model realized in Chalikov and Rainchik (2010).

Third. It is worth while to touch the spectra for wind velocity components. As seen from Fig. 6, the shape of spectrum $S_{u(11)}(k)$ does practically repeat peculiarities of the surface elevation spectrum, $S_{\xi(11)}(k)$, excluding the falling law of the spectrum tail. More weak falling law $S_{u(11)}(k) \propto k^{-2}$ testifies to another nature of velocity fluctuations, differing from one of the waves. More over, having practically the same shape, the vertical component spectrum, $S_{w(11)}(k)$, has intensity of two orders smaller than $S_{u(11)}(k)$. This result is quite clear, if one takes into account the turbulence features corresponding to geometry of the atmospheric boundary layer. Really, in vicinity of wavy surface there is no uniform and isotropic turbulence, spectrum of which is of the form $S(k) \propto k^{-5/3}$ (Monin and Yaglom 1971). Therefore, despite of united nature of fluctuations, one can not expect an isotropic distribution of intensity for the wave-induced velocity fluctuations.

## 7. Conclusion

In conclusion, let us formulate the main results and point out the fields of their possible applicability. In our mind, the most important results are as follows.

---

[7] Theoretical treating of numerous statistical effects in wind waves, provided by their nonlinearity, is rather well represented in a recent book by the author (Polnikov 2007).

7.1. In the through-crests zone of wind waves, beginning at the level of two standard deviations for surface elevation below the mean surface level and ending at the level of three standard deviations above it, the wind profile has a quasi-linear feature in the Cartesian coordinates. This profile is changed by the standard logarithmic profile at the heights above the zone mentioned. (Figs. 1, 4).

7.2. The standard deviations of both horizontal and vertical wind components for wave-induced fluctuations have sharply expressed maxima in the very vicinity of the mean surface level. Intensity of fluctuations for the vertical component of wind velocity is about one order smaller than one for the horizontal component (Fig. 5).

7.3. At the mean surface level the shapes of spectra for both velocity components are identical each to other and have well expressed main and secondary peaks. In our case, the spectra have the tail falling-law proportional to $k^{-2}$, whilst the same law for wind wave spectrum is $k^{-3}$. Spectral intensity of vertical velocity components is in one order smaller than the intensity of horizontal ones.

7.4. Profile of the wave-induced part of momentum flux $\tau_w(z)$ has the well expressed non-monotonic feature with positive and negative extremum, displaying the change of sign in close vicinity of the mean surface level. This feature was named the "sign inversion" effect (Fig. 3). Positive maximum of $\tau_w(z)$ is located at the level of three surface elevation standard deviations above the mean surface level, whilst the negative maximum of twice smaller intensity is located just a little below the mean surface level.

Notice that we do not insist on quantitative values mentioned in the text, they could be refined during further investigations. The most essential result is the fact of presence of the specific features of air flow in the tough–crest zone, concluded here.

The fields of the results application could be as follows.

As it was mentioned in introduction, first of all, the results obtained have a certain scientific interest, extending our understanding the nature of the boundary layer over waves. In this aspect, they could be useful for construction and verification of different numerical models. Examples of such a kind models, dealing with relation between characteristics of wind waves and atmospheric boundary layer, have been represented in our recent paper (Polnikov 2009b).

Another aspect of application of the results obtained deals with the point of quality checking the numerical boundary layer models represented in particular papers (Chalikov and Rainchik 2010, Sulliwan et al, 2000, and others). Taking into account enhanced possibilities of the optical velocimetry technique (described, for example, in Donelan et al. 2004), it could be expected that in the nearest future investigators would be able to measure the wind profile in the trough-crests zone of wind waves with an accuracy sufficient for comparison with the analysis presented in this paper. By this way, the numerical model verification based on the results analogous to the said above becomes real.

Finally, the results presented could play a remarkable role in construction of new, more effective bulk-formulas describing the exchange processes between atmosphere and ocean, as it was mentioned in section 2. More detailed discussion of this issue is out of present paper.

**Acknowledgements**

The author would like to express his sincerely gratitude to Prof. D.V. Chalikov for kind provision the data of numerical simulations, his advices, recommendations, and critical remarks. The work is supported by the grant of the Russian Foundation for Basic Research, # 09-05-00773a.